\documentstyle[12pt,psfig]{article}
\begin{document}
\baselineskip 8truemm
\makeatletter
\makeatother
\begin{titlepage}
\vspace{5mm}
\begin{center}
{\Large {\bf {{Sphaleron in the dilatonic electroweak theory}
}}}\\
\vspace{5mm}
{\bf D.Karczewska\footnote{internet:
dkarcz@usctoux1.cto.us.edu.pl}
 R.Ma\'{n}ka\footnote{internet:
manka@usctoux1.cto.us.edu.pl} }\\
\vspace{5mm}
{\sl Department of Astrophysics and Cosmology,}\\
{\sl University of Silesia, Uniwersytecka 4, 40-007 Katowice,
Poland}\\
\end{center}
\setcounter{equation}{0}
\vspace{5mm}
\centerline{ABSTRACT}
\vspace{5mm}
A numerical study of static, spherically symmetric sphaleron
solutions in the standard model coupled to the dilaton field is presented.
We show that sphaleron is surrounded by strong dilaton cloud
which vanishes inside the sphaleron.\\
\end{titlepage}

\newpage
\noindent
\section{INTRODUCTION}

\noindent
In this paper the electroweak theory is extended by the inclusion of
dilatonic fields.  Such fields appear in a natural way in
Kaluza-Klein theories \cite{tap:1987}, superstring inspired 
theories \cite{ew:1985,sf:1989}  and in
theories based on the noncommutative geometry approach
\cite{abc:1993}.

As previous studies have already shown the inclusion of a dilaton
in a pure Yang-Mills theory has consequences already at 
the classical level. In particular the dilaton Yang-Mills theories
possess `particle - like' solutions with finite energy which are absent in 
pure Yang - Mills case.  On the other hand,
the sphaleron was introduced by Klinkhamer and Manton
\cite{km:1983} 
to describe a static electroweak gauge field configuration that 
constitutes a saddle point between two vacua differing by non
trivial topology (the hedgehog topology).  Analogous equations
have recently been obtained for the t'Hooft-Polyakov monopole model   
coupled to dilatonic field \cite{fg:1996}.
There is also growing interest
 \cite{krs:1985} in baryon number violation within the standard model
induced by sphalerons. The rate of baryon number violating processes
depends on the energy of sphaleron \cite{am:1988}. 

The aim of
this paper is to examine the properties of the sphaleron solution in the
presence of dilatonic field in the electroweak theory. We demonstrate the
existence of  
spherically symmetric dilatonic cloud  surrounding the sphaleron. 
We also discuss very interesting properties of
the Higgs field in the dilaton electroweak theory.

\noindent
\section{THE DILATONIC ELECTROWEAK THEORY}

\noindent
One of the interesting features of the standard model is the
scale
invariance of highly symmetric phase. 
Quantum effects anomalously cause its break up and produce nonvanishing 
cosmological
constant. The classical scale invariance offers a link between the 
standard model and gravity, which is successfully implemented in
the Jordan-Brans-Dicke
theory \cite{jbs:1959} of scalar-tensor theory of gravity.

In this paper we consider the
Glashow-Weinberg-Salam dilatonic model with $SU_{L}(2)\times
U_{Y}(1)$
symmetry
\begin{eqnarray}
{\cal L} &=&  -\frac{1}{4} e^{2  \varphi (x)/f_0 } F^{a}_{\mu
\nu}F^{a\mu \nu} -
\frac{ 1}{4 } e^{2  \varphi (x)/f_0 } B_{\mu \nu}B^{\mu \nu}
\nonumber\\
& + & \frac{ 1}{4 }{\partial}_{\mu} \varphi {\partial}^{\mu}
\varphi + (D_{\mu}H)^{+}D^{\mu}H -U(H) e^{-2 \varphi(x)/f_0}
\end{eqnarray}
with the $SU_{L}(2)$ field strength tensor $F^{a}_{\mu \nu}
= \partial_{\mu}W^{a}_{\nu} - \partial_{\nu}W^{a}_{\mu} +
g\epsilon_{abc}W^{b}_{\mu}W^{c\nu}$ and the $U_{Y}(1)$ field tensor
 $B_{\mu \nu} = \partial_{\mu}B_{\nu} - \partial_{\nu}B_{\mu}$.
The covariant derivative is given by
$D_{\mu} = \partial _{\mu} -\frac{1}{2}igW^{a}_{\mu}\sigma^{a}-
\frac{1}{2}g^{'}YB_{\mu}$, where $B_{\mu}$ and 
$W_{\mu}=\frac{ 1}{2 } W^{a}_{\mu}\sigma^{ a}$
are local gauge fields associated with $U_{Y}(1)$ and $SU_L(2)$
symmetry groups respectively. $Y$ denotes the  hypercharge. The
gauge group is a simple product of $U_{Y}(1)$ and $SU_{L}(2)$
hence we have two gauge couplings $g$ and $g'$.
The generators of gauge groups are: a unit matrix for $U_{Y}(1)$ and
Pauli matrices for $SU_{L}(2)$. In the simplest version of the
standard model a doublet of Higgs fields is introduced 
$H = \left( \begin{array}{c}
H^{+} \\ H^{0} 
\end{array}   \right)$,
with the Higgs potential
\begin{equation}
U(H) = \lambda \left( H^{+}H - \frac{1}{2}v^{2}_0 
 \right)^{2} 
\end{equation}
The $f_0$ and $v_0$ parameters in the Lagrangian function (1) 
determine the dilaton scale   $f_0 = 10^7 GeV$ \cite{ro:1994}, and the electroweak 
symmetry breaking scale $v_0 = 250 GeV $. 
The form of the potential $(2)$ leads to vacuum degeneracy, 
nonvanishing vacuum expectation value of the
Higgs field and consequently to fermion and boson
masses. In this process of spontaneous symmetry breaking the Higgs field 
acquires nonzero mass.\\

The Euler-Lagrange equations for the lagrangian (1) are scale-invariant: 
$ x^{\mu} \rightarrow x'^{\mu} = e^{\frac{u}{f_0}} x^{\mu}$,\\
$\varphi \rightarrow \varphi' = \varphi +u $,\\
$H \rightarrow H' = H $,\\
$W_{\mu}^a \rightarrow W_{\mu}^{,a} = W_{\mu}^a $,\\
$B_{\mu} \rightarrow B'_{\mu} = B_{\mu} $.\\
These transformations change the Lagrange function in a following way 
\begin{equation}
L \rightarrow L' = e^{- \frac{2u}{f_0}} L
\end{equation}

\noindent
\section{THE DILATONIC SPHALERON}

\noindent
Let us now consider the sphaleron type solution in the electroweak theory
with dilatons. The sphaleron may be interpreted as inhomogeneous bosons
condensate $ \varphi(x),W_\mu ^a(x)$. Let us assume for simplicity that $
g'=0$. (In \cite{kb:1991} the sphaleron theory was also considered also 
for $ g' \ne 0 $.)  Firstly, we make the following anzatz 
for the sphaleron Higgs field 
\begin{equation}
H=\frac 1{\sqrt{2}}v_0 U(x)h(r)\left( 
\begin{array}{c}
0 \\ 
1
\end{array}
\right) 
\end{equation}
where $U(x)=i\sum \sigma ^an^a$ and $n^a=\frac{r^a}{r}$ describe the
{\it hedgehog} structure. This produces a nontrivial topological charge of the
sphaleron. The topological charge is equal to the Chern-Simons number. Such
a {\it hedgehog} structure determines the asymptotic shape of the
sphaleron with gauge fields different from zero 
\begin{equation}
W_i^a=\epsilon _{aij}n^j\frac{1-f(r)}{gr},\\
\varphi(x)=f_0 s(r).
\end{equation} 
It is convenient to introduce the dimmentionless variable $x$ defined 
as $x=M_Wr={r}/{r_W}$,
where $M_W^2=\frac 14g^2v_0^2\sim 80GeV$, $r_W=\frac 1{M_W}\sim 10^{-18}cm$. 
Spherical symmetry is assumed for the dilaton field $s(r)$, as well as 
for the Higgs field $ h(r)$ and the gauge field $f(r)$, leading 
to the following expression for the total energy
\begin{equation}
E= \frac{4\pi M_W}{g^2} \int \rho_0(x)x^2dx,
\end{equation}
where the energy density is:
\begin{eqnarray}
\rho_0(x) &=& 2h'(x)^2 + \frac{1}{\alpha} s'(x)^2 \nonumber\\
&+& \frac{1}{x^2} e^{2s(x)} \left\{f'(x)^2 + \frac{1}{2x^2}(f(x)-1)^2
(f(x)-3)^2 \right\} \nonumber\\
&+& \varepsilon \left( h(x)^2 -1 \right)^2 e^{-2s(x)}
+ \frac{1}{x^2}(f(x)-3)^2 h(x)^2.
\end{eqnarray}
$M_H^2=2\lambda v_0^2$ determines the Higgs mass;
$ \alpha = \frac{ M_w^2}{f_0^2 g^2}$, and
$\varepsilon = \frac{M_H^2}{2M_W^2}$ are dimmensionless 
parameters which completly determine
the sphaleron system.

The resulting Euler-Lagrange equations are following, where the $s(x)$ 
function describes the gauge field inside the sphaleron,
\begin{eqnarray}
&f^{\prime \prime}(x)&+ 2f'(x)s'(x) + (3-f(x))h(x)^2 e^{-2s(x)} \nonumber\\ 
&-& \frac{1}{x^2}(f(x)-1)(f(x)-2)(f(x)-3) = 0.
\end{eqnarray}
$h(x)$ function describes the Higgs field in our theory, it satisfies
the following equation: 
\begin{eqnarray}
&h^{\prime \prime}(x)&+ \frac{2}{x}h'(x) + \varepsilon e^{-2s(x)}
(1-h^2(x))h(x) \nonumber \\
&-&\frac{1}{2x^2}(3 - f(x))^2h(x)=0.
\end{eqnarray}
The $s(x)$ function describing the dependence of a dilaton field 
on $x$ in the extended electroweak theory obeys the equation:
\begin{eqnarray}
&s^{\prime \prime}(x)&+ \frac{2}{x}s^{\prime}(x) + \alpha e^{2s(x)} \left\{-
\frac{2}{x^2}f'(x)^2 - \frac{1}{x^4}(f(x)-1)^2(f(x)-3)^2 \right\} \nonumber\\
&+& 2\varepsilon \alpha e^{-2s(x)}(1-h^2(x))^2 = 0,
\end{eqnarray}
with a dimensionless constant $ \alpha \sim
(\frac{M_W}{gf_0})^2 \sim 10^{-9}$. This means that the 
dilaton field is essentially a free field. The simplest solutions
are the global ones corresponding to the vacuum with broken symmetry in
the standard model.  It is obvious that far from the center of the
sphaleron our solutions should describe the normal broken phase which is
very well known from the standard model.  From the known asymptotic solutions 
for $x \rightarrow 0$:
\begin{equation}
f(x) = 1 + 2 t^2 x^2 + O(x^3),
\end{equation}
\begin{equation}
h(x) = ux + O(x^3),
\end{equation}
\begin{equation}
s(x) = s_0 - \frac{1}{2} s_0 v^2 x^2 + O(x^3), 
\end{equation}
and for $x \rightarrow \infty$:
\begin{equation}
f(x) = 3 - f_{\infty} e^{-x},
\end{equation}
\begin{equation}
h(x) = 1 - \frac{h_{\infty}}{x} e^{- \sqrt{2 \epsilon}},
\end{equation}
\begin{equation}
s(x) = s_{\infty} - \frac{d_{\infty}}{x}, 
\end{equation}
we are able to construct a four-parameter family of trial solutions: 
\begin{equation}
f(x)=1+2\tanh^2(tx),
\end{equation}
\begin{equation}
h(x)=\tanh(ux),
\end{equation}
\begin{equation}
s(x)=s_0 e^{-\frac{1}{2} v^2 x^2},
\end{equation}
where $t,u,v,s_0$ are parameters to be determined by the variational 
procedure. Parameters $ u,t,v$ may be determined in this way to satisfy
boundary conditions (12), (13), (14).  For
parameter $s_0$, by minimizing the sphaleron mass (6) with respect 
to $s_0$ we may choose the trial
sphaleron configuration with the minimal energy (mass). 
In so doing, the trial solutions
have a proper behaviour for $ x \rightarrow 0 $ and describes the sphaleron 
configuration with minimal energy. Then this trial function can serve as
an initial solutions for solving the coupled systems of differential
equation (9), (10), (11) by the use of
{\it Mathematica} and {\it Scientific Workplace} with
Maple library. 

Specifically, we substituted trial functions (18), (19), (20)
into the differential equations (9), (10), (11), then performed Taylor series 
expansion for 
$ x \rightarrow 0 $ and solved system of algebraic equation for $u,t$
and $v$,
\begin{equation}
u = \sqrt{\frac{3}{10}} \sqrt{\varepsilon + \frac{6}{5}(3+ 
\sqrt{9+5\varepsilon})} e^{-s_0}
\end{equation}
\begin{equation}
t=\sqrt{ \frac{3}{10}}e^{-s_0} \sqrt{3 + \sqrt{9 + 5 \varepsilon}}
\end{equation}
\begin{equation}
v= \sqrt{\frac{2}{3}} e^{-s_0} \sqrt{ \frac{\alpha}{s_0}(\varepsilon - 
\frac{54}{25}(3 + \sqrt{9+5\varepsilon})} 
\end{equation}

The numerical solutions are close to the trial 
minimal functions (18), (19), (20), which are presented on Figs. 2-4.
The relevant values of the parameters are those which 
minimize the energy, Fig.1. For
example, with the standard values of $M_W=80.6GeV$, $M_Z=91.16GeV$, $%
M_H=350GeV$ we found the numeric solutions $t,u,k$, as functions depending
on the initial conditions of the dilaton field $s(0)=a$ in the center of the
sphaleron. 
Our solutions describe both the behavior of Higgs field and
gauge field inside the sphaleron and the shape of the dilaton cloud
surrounding the sphaleron. Such a cloud is large and extends far outside the
sphaleron \cite{km:1996}. 

The sphalerons might be created during the first order phase 
transition in the expanding universe as inhomogeneous solutions 
of the equations of motion.  The bubbles left after the phase 
transition probably took place in the early universe, break the 
CP and C symmetry on their walls and can cause the breaking 
of baryonic symmetry. Detailed consideration of this 
problem will be the subject of a separate paper.

\noindent
\section{CONCLUSIONS}

\noindent
The aim of this paper was to present a numerical study of the 
classical sphaleron solutions of the SU(2) Yang - Mills theory 
coupled to the dilaton fields.

Dilaton field appears naturally in low energy sector of the effective
field theories derived from superstring theories or noncommutative field
theory approach. Numerical solutions suggest that sphaleron possess
an `onionlike' structure. In the small inner core the 
scalar field is decreasing with global gauge symmetry 
restoration $ SU(2) \times U(1) $.
In the middle layer the gauge field undergoes sudden change.
It is very interesting that sphaleron coupled to dilaton field
has also an outer shell, where dilaton field changes drastically.
The spherically symmetric dilaton solutions coupled to the gauge field 
or gravity are interesting in their own and may moreover 
influence the monopole catalysis of baryogenesis induced by sphaleron.

We would like to thank Dr Marek Biesiada for 
fruitful discussion and Dr Tien D. Kieu for interesting comments.\\
This paper is sponsored by the Grant KBN 2 P304 022 06.

\newpage
{\bf FIGURE CAPTIONS}\\

Figure 1.  The sphaleron energy $E_o$ as a function of the variarional\\ 
           parameter $s_o$.\\

Figure 2.  The dependence of the Higgs field $h(x)$ on $x$.\\

Figure 3.  The dependence of the gauge field $f(x)$ on $x$.\\

Figure 4.  The dependence of the dilaton field $s(x)$ on $x$.\\

\begin{figure}
\psfull
\psfig{file=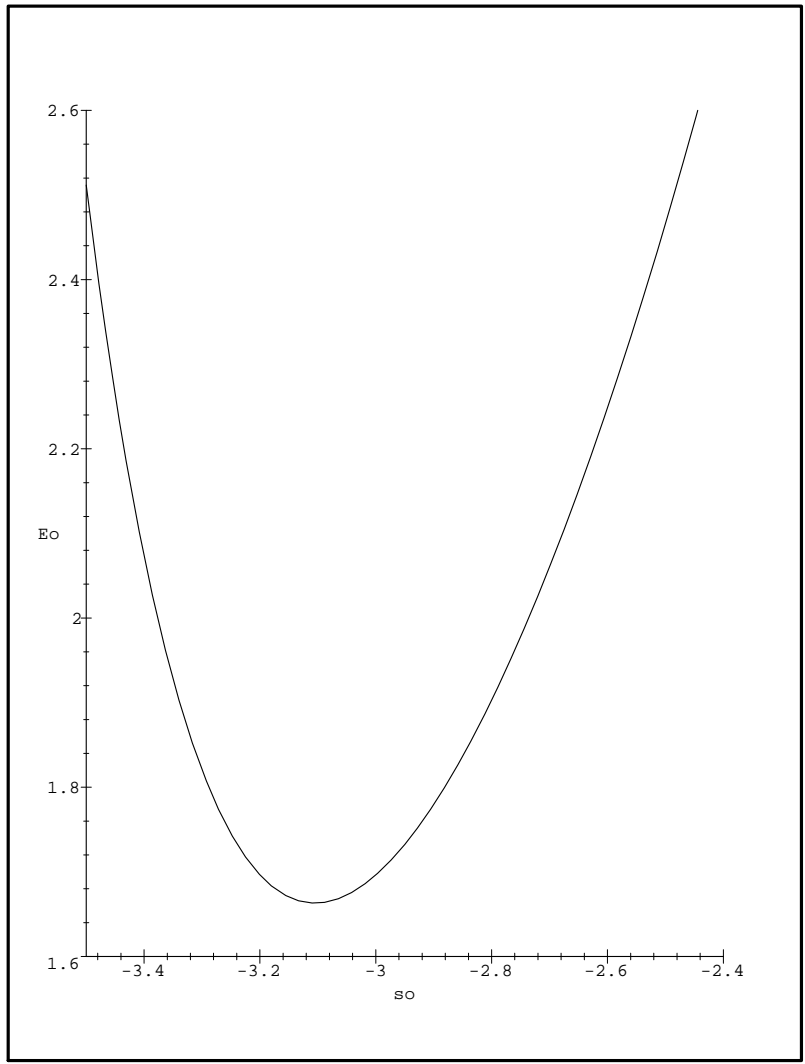,width=12cm}
\caption{The sphaleron energy $E_o$ as a function of the variational
parameter $s_o$ }
\end{figure}

\begin{figure}
\psfull
\psfig{file=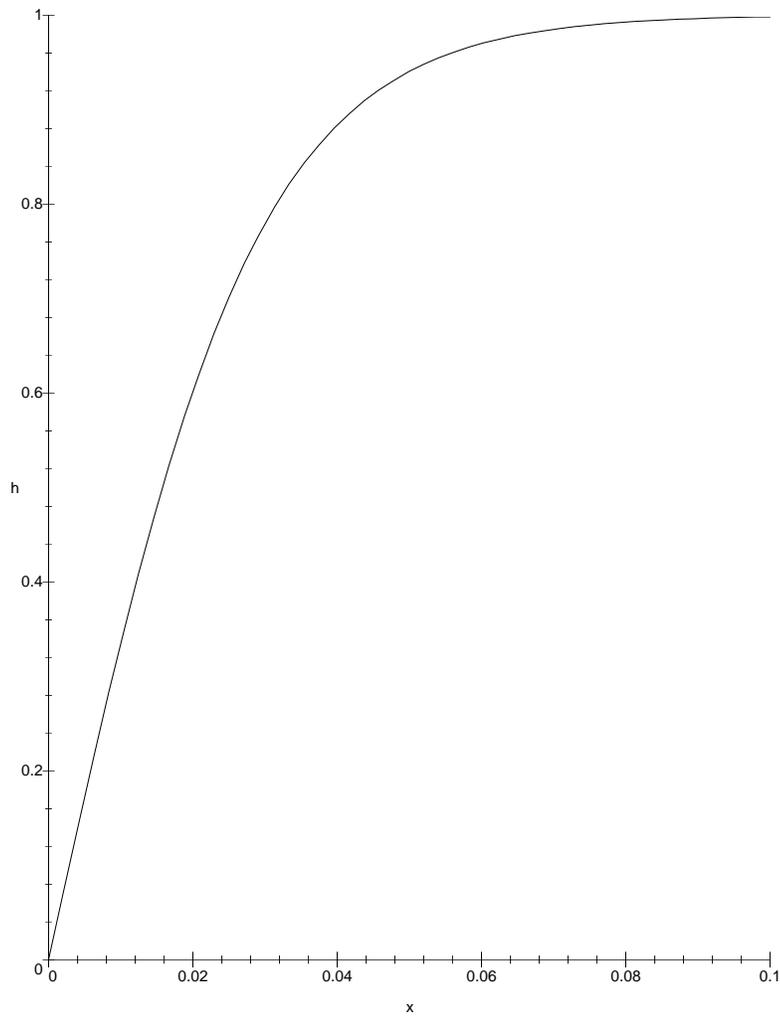,width=12cm}
\caption{The dependence of the Higgs field $h(x)$ on $x$.}
\end{figure}

\begin{figure}
\psfull
\psfig{file=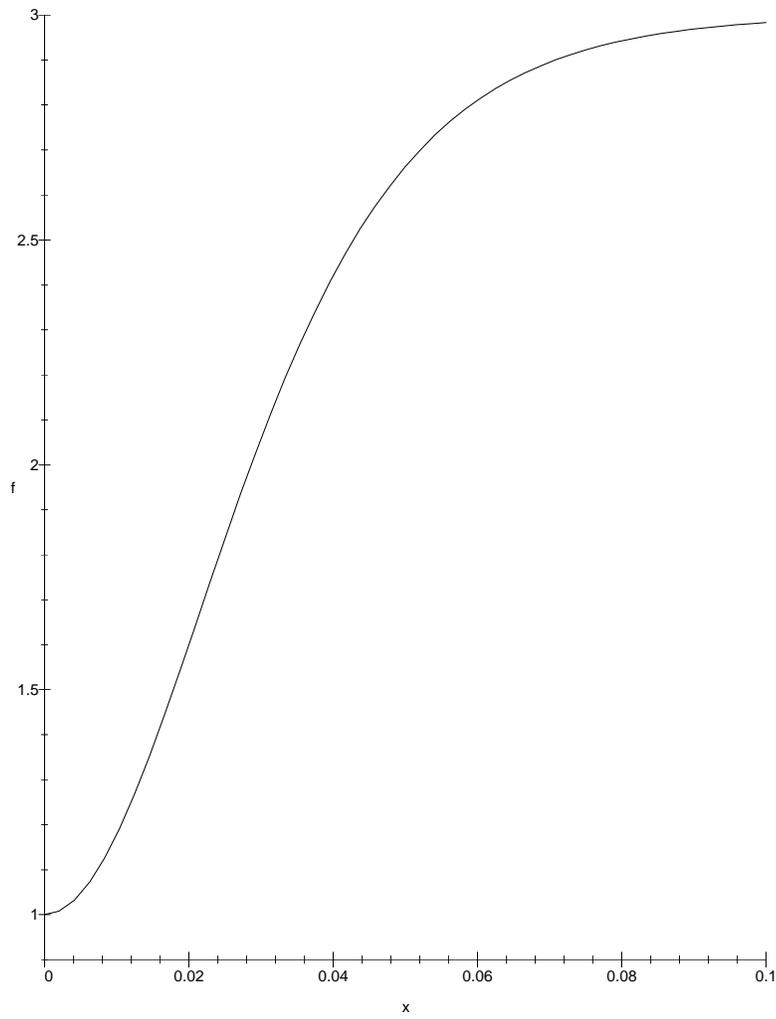,width=12cm}
\caption{The dependence of the gauge field $f(x)$ on $x$.}
\end{figure}

\begin{figure}     
\psfull
\psfig{file=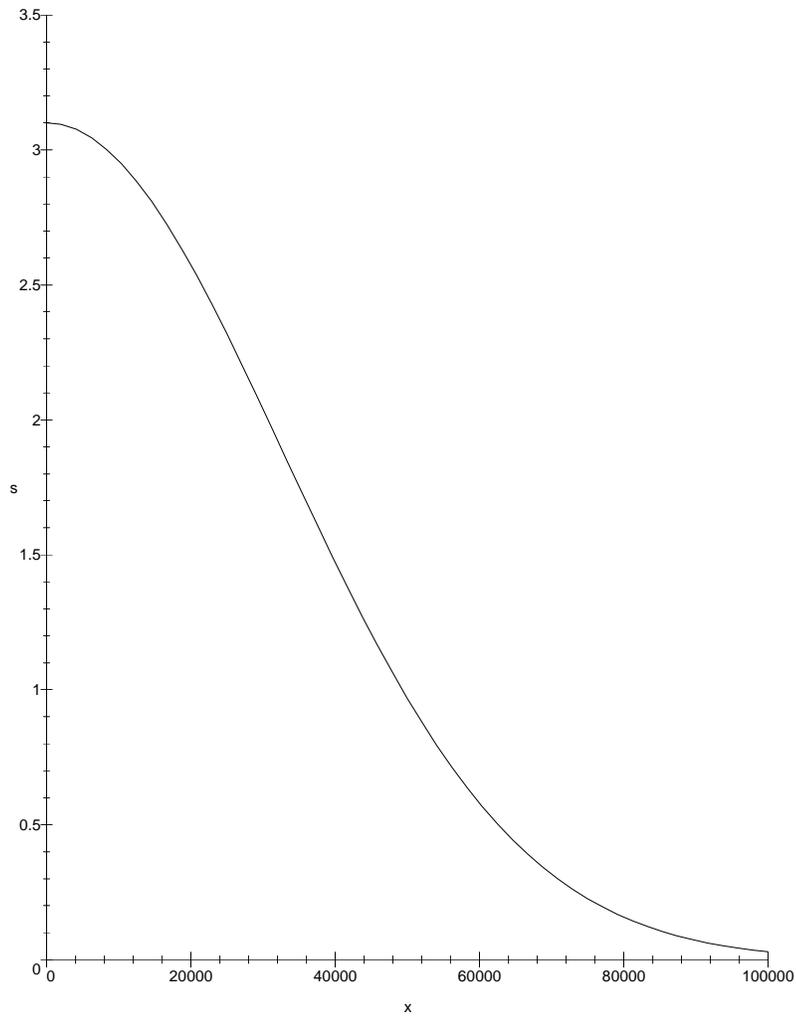,width=12cm}
\caption{The dependence of the dilaton field $s(x)$ on $x$.}
\end{figure}

\end{document}